\documentclass[11pt]{article}

\usepackage{amsmath,amsfonts,amssymb}
\usepackage{graphicx,color}
\usepackage{graphics}

\usepackage{epsf}

\newcommand{\be}{\begin{equation}}
\newcommand{\ee}{\end{equation}}
\newcommand{\ba}{\begin{eqnarray}}
\newcommand{\ea}{\end{eqnarray}}
\newcommand{\nn}{{\nonumber}}
\newcommand{\bc}{\begin{center}}
\newcommand{\ec}{\end{center}}

\begin{document}
\begin{titlepage}
\begin{flushright}
{\bf March 8, 2007} \\
DAMTP-07-15\\
hep-th/0703068 \\
\end{flushright}
\begin{centering}
\vspace{.2in}

{\large {\bf Geometry and topology of bubble solutions from gauge
theory}}

\vspace{.3in}

Heng-Yu Chen${}^{1}$, Diego H. Correa${}^{1}$ and Guillermo A. Silva${}^{2}$ \\
\vspace{.2 in}
${}^{1}$DAMTP, Centre for Mathematical Sciences \\
University of Cambridge, Wilberforce Road \\
Cambridge CB3 0WA, UK \\
\vspace{.2in}
and \\
%\vspace{.2in}
\vspace{.1 in}
${}^{2}$Departmento de F\'{\i}sica  \\
Universidad Nacional de La Plata\\
CC 67, 1900,  La Plata, Argentina\\
\vspace{.4in}

{\bf Abstract} \\
\end{centering}

We study how geometrical and topological aspects of certain
$\frac12$-BPS type IIB supergravity solutions are captured by the
${\cal N}=4$ Super Yang-Mills gauge theory in the AdS/CFT context.
The type IIB solutions are completely characterized by arbitrary
droplets in a plane and we consider, in particular, concentric
droplets. We probe the dual $\frac12$-BPS operators of the gauge
theory with  single traces and extract their one-loop anomalous
dimensions. The action of the one-loop dilatation operator can be
reformulated as the Hamiltonian of a bosonic lattice. The operators
defining the Hamiltonian encode the topology of the droplet. The
axial symmetry of the droplets turns out to be essential for
obtaining the spectrum of the Hamiltonians. In appropriate BMN
limits, the near-BPS spectrum reproduces the spectrum of near-BPS
string excitations propagating along each individual edge of the
droplet of the dual geometric background. We also study
semiclassical regimes for the Hamiltonians. We show that for
droplets having disconnected constituents, the Hamiltonian admits
different complimentary semiclassical descriptions, each one
replicating the semiclassical description for closed strings
extending in each of the constituents.

%\vspace{.05in}
%\baselineskip=.3in
\end{titlepage}

%%%%%%%%%%%%%%%%%%%%%%%%%%%%%%%%%%%%%%%%%%%%%%%%%%%%%%%%%%%%%%%%%%%%%%%%%%%%%%%%%%%%%%%%%%%%
%%%%%%%%%%%%%%%%%%%%%%%%%%%%%%%%%%%%%%%%%%%%%%%%%%%%%%%%%%%%%%%%%%%%%%%%%%%%%%%%%%%%%%%%%%%%
\section{\label{intro}Introduction}

%\section{Introduction}
%\label{intro}

The notion of emergent geometry in the AdS/CFT correspondence
\cite{adscft} holds certain appeal. The general idea is that
judicial choices of BPS operators in the gauge theory are capable of
encoding the characteristics of the corresponding dual geometries
\cite{Btoy,LLM}.

A paradigmatic example was provided by \cite{Btoy},  focusing on the
$\frac{1}{2}$-BPS operators of $\mathcal{N}=4$ SYM  constructed
exclusively  from a complex adjoint chiral scalar $Z$. The dynamics
of this sub-sector can be described by gauged  harmonic oscillators,
which in turn can be mapped into a Fermi problem. The ground state
of the system corresponds to a completely occupied Fermi sea forming
a disc in the phase space. This distribution, known as the droplet,
can then be associated, at the semi-classical level, with the
allowed eigenvalues of the scalar field $Z$  \cite{Btoy}. Other
$\frac{1}{2}$-BPS `excitations' correspond to disc distortions,
further splitting into isolated droplets and creation of holes
within these droplets. All these lead to  droplets with non-trivial
shapes and topologies \cite{LLM,Horava,BBPS,Brown}. This class of
$\frac12$-BPS  states preserve 16 supersymmetries as well as a
bosonic $R\times SO(4)\times SO(4)$ symmetry group. The dual type
IIB supergravity backgrounds, known as Lin-Lunin-Maldacena (LLM)
geometries, were constructed by fibering two $S^{3}$ over a two
dimensional plane \cite{LLM}. The base of the fibration in this
construction was proposed to be identified with the droplets
describing the eigenvalue distribution in the gauge theory. In
particular, the $AdS_{5}\times S^{5}$ geometry is shown to
correspond to a circular disc droplet, whereas single particle and
hole creations were identified with the nucleations of a giant
graviton in $AdS_{5}$ and $S^{5}$ respectively \cite{Btoy,LLM}. The
bottom line of the construction is that when the number of giant
gravitons becomes large, the back-reacted non-singular geometry gets
parameterized in terms of a single scalar function obeying a
Laplace-like differential equation whose boundary conditions on a
two dimensional plane can only take the values $\pm \frac12$.

Generically, a probe closed string propagating on a LLM geometry
should be seen in the dual gauge field theory  as a non-BPS operator
given by a single trace operator describing the excitations (probe
string) on top of a {$\frac12$}-BPS operator (LLM geometry). The
study of the one-loop anomalous dimensions for such gauge theory
operators turns out to be very rewarding. For the case of simply
connected droplets, the gauge theory Hamiltonian describing probe
strings  was recently derived  \cite{sam}. This Hamiltonian can be
written in terms of certain operators obeying an algebra that
encodes, in a simple fashion, the moments describing the geometry of
the droplet. Moreover, using the coherent state basis for the
operators, a semi-classical action was derived, from which it was
possible to reconstruct the region of the LLM droplet plane where
the dual string propagates \cite{sam}. The approach described in
\cite{sam,BCV,BCV2} is similar in spirit to the coherent state
approach in the context of spin-chain/spinning string correspondence
\cite{Kruczenski,kru}, where the coherent state action for the
integrable spin chain Hamiltonian \cite{minahan} was identified with
the spinning-string action in certain ``fast-string'' limit.

Making the existing picture more precise is the focus of our paper.
In particular, we will obtain quantitative information about the
dual metric, for droplets of general shapes and topologies, from
gauge theory computations. For definiteness we will consider non-BPS
excitations, or string probe states, in topologically non-trivial
LLM backgrounds, evaluate their scaling dimensions and importantly,
distinguish them from their counterparts in $AdS_{5}\times S^{5}$.
From the dual gauge theory perspective, the key steps towards these
aims amount to defining the appropriate field theory operators and
compute the action of the dilatation operator whose eigenvalues give
the scaling dimensions.

The generalization we carry out is interesting since the geometric
backgrounds we consider are topologically different from
$AdS_5\times S^5$ or any LLM geometries associated with simply
connected droplets. In this regard, it is important to understand
how topology is revealed from purely gauge theory means. In
particular, the appearance of additional edges for the droplet
indicates the existence of new disconnected sets of null geodesics
in the geometry. Let us recall that for the case of circular disc,
BPS and near-BPS string excitations are represented by single traces
having a large number $J$ of complex chiral scalar fields $Z$
\cite{BMN}, furthermore these excitations get localized at the
unique edge of the circular droplet \cite{LLM}. When droplets with
multiple edges are considered, one should identify new types of
excitations and argue how they get localized at each individual
edge.  Another appealing situation to consider are the non-connected
droplets, since one should then identify within the gauge theory the
operators dual to strings whose propagations are restricted to each
disconnected droplet.

The simplest setup for studying multiple edges and disconnected
droplets is that of concentric droplets. An advantageous feature of
these axially symmetric droplets is that the dual LLM geometry
defining functions are explicitly given as superposition of uniform
discs.  The one-loop Hamiltonian we will derive on the gauge theory
side corresponding to strings probing concentric LLM geometries, will
 be seen as the Hamiltonian of a bosonic lattice along the lines
of \cite{sam,BCV,BCV2}. Interestingly, the axial symmetry of the
droplets will be essential for finding the Hamiltonian spectrum.

The rest of the paper is organized as follows: In section \ref{bps}
we present the $\frac{1}{2}$-BPS operators we will be working with:
they are characterized by a concentric distribution of eigenvalues
and as should become clear dual to concentric LLM geometries. In
section \ref{string} we excite them with single traces. We focus on
two particular cases: the annular droplet which is connected but has
two edges and a droplet consisting of a disc and a disconnected
annulus. We then derive the Hamiltonians corresponding to the
one-loop anomalous dimension operators for non-BPS excited states.
By taking appropriate BMN-like limits, we show the spectra of these
Hamiltonians precisely match with the spectra of the corresponding
dual string states. Furthermore, by considering the semi-classical
coherent state action for these Hamiltonians, we show the agreement
with the Polyakov action of the dual string solution in certain
fast-string limit. We summarize and discuss potential research
directions in section 4. In an appendix, we include details on LLM
geometries as well as their BMN limits; we also list some
semi-classical string solutions in backgrounds constructed from
concentric droplets.

%%%%%%%%%%%%%%%%%%%%%%%%%%%%%%%%%%%%%%%%%%%%%%%%%%%%%%%%%%%%%%%%%%%%%%%%%%%%%%%%%%%%%%%%%%%%
%%%%%%%%%%%%%%%%%%%%%%%%%%%%%%%%%%%%%%%%%%%%%%%%%%%%%%%%%%%%%%%%%%%%%%%%%%%%%%%%%%%%%%%%%%%%
\section{Concentric droplets}
\label{bps}

Half-BPS states of $U(N)$ ${\cal N}=4$ SYM can be described in terms
of a gauged $N\times N$ normal matrix model with a harmonic
oscillator potential  \cite{Btoy,CJR}. Each supersymmetric state is
associated to a wave-function of the normal matrix model\footnote{
The wave function depends  on the $N$ complex eigenvalues of the
normal matrix model.}. The normalization of each BPS state coincides
with the partition function of a random matrix model for a
particular ensemble $W$,
\begin{equation} 
\int \prod_{i=1}^N d^2z_i\, |\psi(z)|^2 =  \int \prod_{i=1}^N
d^2z_i\, \exp\left(\sum_{j}W(z_j,\bar z_j)+2\sum_{j<k}\log|z_j-z_k|
\right)\, . 
\label{partition} 
\end{equation} 
The logarithm in the exponential is attributed to the Van Der Monde
determinant, which arises from writing the integration measure in
terms of the eigenvalues of the matrices (see \cite{zabrodin} for a
review) and is the origin of the system becoming fermionic. One
effectively ends up with a system consisting on $N$ one-dimensional
fermions.

The gauge theory ground state, which is dual to  the $AdS_5\times
S^5$  geometry, corresponds to the gaussian ensemble, i.e.
\begin{equation}
W(z,\bar z)= -|z|^2\, . \label{gaussian} \end{equation}
In the large $N$ limit, the semi-classical continuum approximation
to the matrix model partition function (\ref{partition})  leads to a
characterization of the states  in terms of a density distribution
of eigenvalues $\rho$ defined on a two dimensional plane. The
support of the density distribution determines what we call the
droplets. For the gaussian potential (\ref{gaussian}), the saddle
point analysis shows that the partition function (\ref{partition})
is dominated by a uniform distribution of eigenvalues on a disc of
radius $\sqrt{N}$  centered at the origin \cite{zabrodin}.

Arbitrary $\frac12$-BPS states are constructed as `excitations'
above the ground state and can be represented as \cite{Btoy,BBPS}
\begin{equation}
|\psi\rangle = \exp\left({\rm
tr}\left(\Omega(Z)\right)\right)|\psi_0\rangle\, . \label{exci}
\end{equation}
The corresponding normalization is again a  matrix model partition
function, but with the potential ensemble $W$ now generalizes to:
\begin{equation}
W(z,\bar z)= -|z|^2+\Omega(z)+\overline{\Omega(z)}\, .
\label{potential}
\end{equation}

A harmonic $\Omega(z)$  function guarantees that in the large $N$
limit, the system continues to be dominated by a uniform
distribution of eigenvalues, with the droplet shape depending on the
potential ensemble $W$. BPS states corresponding to polynomial
potentials and their string excitations have recently been studied
in \cite{sam}.

Exciting the ground state with the operator $\det(Z-\lambda)$
induces a potential $\Omega(z)=\log(z-\lambda)$, which generates a
hole in the circular droplet at complex position $\lambda$
\cite{Btoy,BBPS}. This corresponds  on the string side to   a single
non-back-reacting spherical giant graviton in the $AdS$ bulk. In
order to  obtain a hole of size comparable to the area of the
droplet and create a back-reacted geometry,  the number of giant
gravitons needs to be comparable to $N$. Hence we need to excite the
ground state with a product of $M\simeq{\cal O}(N)$ determinants. We
will thus consider a density distribution of holes $\sigma(\lambda)$
giving the potential
\begin{equation}
\Omega(z)=\int d^2\lambda\ \sigma(\lambda)\log(z-\lambda)\, .
\label{oman}
\end{equation}
An important example is the following: Take  $\sigma(\lambda)$ to be
$\frac1\pi$ in the interior of a disc $D$ of radius $R_1= \sqrt{M}$
centered at the origin and zero elsewhere, the potential $W$ takes
the form
\begin{equation}
W(z,\bar z)= -|z|^2 + \frac{2}{\pi}\int_{D} d^2\lambda\
\log|z-\lambda|\, . \label{poan}
\end{equation}
The second term in (\ref{poan}) is nothing but the electrostatic
potential in a two-dimensional plane due to an uniformly charged
disc. Performing the integral, the  potential (\ref{poan}) reads
\begin{equation}
W(z,\bar z)= \left\{
\begin{array}{c l}
0&  {\rm if}\ |z|\leq R_1\\
-|z|^2 + R_1^2(1+2\log(|z|/R_1)) & {\rm if}\ |z|> R_1
\end{array}\right.
\label{poan2}
\end{equation}
The random matrix problem with potential (\ref{poan2}) is tractable
even for finite $N$, we will nevertheless  be interested in the
large $N$ limit later.

\begin{figure}%[htb]
\begin{center}
\epsfxsize=6in \leavevmode \epsfbox{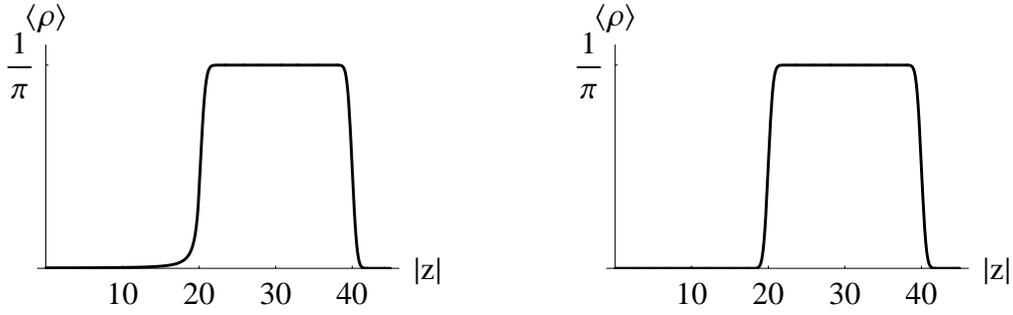}
\end{center}
\caption{Eigenvalue distributions for $N=1200$ and $M = 400$ for:
(left) homogeneously  distributed holes on a disc and (right) all
holes placed at the origin.} \label{fig}
\end{figure}

Let $\rho(z)$ be the eigenvalues distribution, its mean value in the
matrix model (\ref{partition}) is given by (see \cite{zabrodin} for
a review)
\begin{equation}
\langle \rho(z) \rangle_N = e^{W(|z|)}\sum_{i=0}^{N-1}
\frac{|z|^{2i}}{h_i}\,, \label{den}
\end{equation}
and for the axially symmetric potentials, the constants $h_i$ are
given by
\begin{equation}
\int d^2z\ z^i\overline{z}^j e^{W(z,\bar z)} = h_i \delta_{ij}
\,. \label{orto}
\end{equation}
The potential (\ref{poan2}) gives
\begin{equation}
h_i= \frac{R_1^{2i+2}}{i+1} +
\left(\frac{e}{R_1^2}\right)^{R_1^2}\Gamma(n+1+R_1^2,R_1^2) \,.
\label{hache}
\end{equation}
Here $\Gamma(a,x)$ is the incomplete gamma function. Plotting
(\ref{den}) for different values of $N$ and $R_1$, it is easy to see
that, as $N$ and $R_1^2$ are taken large, $\langle \rho(z)\rangle_N$
approaches $1/\pi$ for $R_1<|z|<\sqrt{R_1^2+N}$ or zero otherwise
(see Figure \ref{fig}).

Let us analyze the continuum (large $N$) limit for an arbitrary
distribution of holes $\sigma(\lambda)$.  In other words, consider a
state obtained by the product of $M\sim {\mathcal{O}}(N)$
determinants  at arbitrary positions $\lambda_\alpha$ exciting the
ground state,
\begin{equation}
|\psi\rangle =
\prod_{\alpha=1}^M\det(Z-\lambda_\alpha)|\psi_0\rangle\, .
\label{excid}
\end{equation}
The integrand in (\ref{partition}) then takes the form
\begin{eqnarray}
|\psi|^2 &\! =\!& \exp\left(-\int\! d^2\!z\,|z|^2 \rho(z)
+2\int\!\!\int\!d^2\!z\,d^2\!\lambda\,\rho(z)\sigma(\lambda)\log|z-\lambda|
\right. \nn \\
&& \qquad\qquad+\left.\int\!\!\int\!d^2\!z\, d^2\!{z'}
\rho(z)\rho(z')\log|z-z'| \right)\, , \label{nor}
\end{eqnarray}
where the eigenvalue and hole density distributions $\rho(z)$ and
$\sigma(z)$ are constrained to satisfy
\begin{equation}
\int\! d^2\!z\, \rho(z) = N\, , \qquad \int\! d^2\!z\, \sigma(z)
= M\, .
\label{cons}
\end{equation}
The normalization (\ref{nor}) is dominated by the eigenvalue
distribution $\rho(z)$ maximizing the exponential. Taking a
variation with respect to $\rho(z)$, one is led to the following
equation of motion,
\begin{equation}
 -|z|^2  +2\! \int\!d^2\!\lambda\,\sigma(\lambda)\log|z-\lambda|
+2\!\int\! d^2\!{z'} \rho(z')\log|z-z'|=0 \, . \label{eom}
\end{equation}
Now, acting on (\ref{eom}) with the two-dimensional Laplace operator
we get a consistency equation:
\begin{equation}
 -4  + 4\pi \sigma(z) + 4\pi\rho(z)=0 \, .
 \label{eom2}
\end{equation}
This equation justifies the selection $\sigma(z)=\frac1\pi$ in
(\ref{poan}). Choosing the hole density $\sigma(z)$ to be
exclusively $1/\pi$ or zero guarantees having no partially filled
regions. The droplet density $\rho(z)$ vanishes in the region where
the hole density $\sigma(z)$ takes the value $1/\pi$, and takes the
value $1/\pi$ where $\sigma(z)$ vanishes. This choice is the dual
version of the one appearing on the gravity side ensuring singularity
free geometries \cite{LLM}.

In the example  discussed above (equations
(\ref{poan})-(\ref{poan2})),  $\sigma(z)$ was taken to be constant
over a disc of radius $R_1$, and as result the eigenvalues got
distributed in an annular domain with radii $R_1=\sqrt{M}$ and
$R_2=\sqrt{M+N}$.  However, the potential (\ref{poan}) leading to
the uniform annular eigenvalue distribution is not unique in the
large $N$ limit. Alternatively, consider exciting the ground state
with $\det(Z)^M$, which corresponds to placing all the holes at the
origin\footnote{The same operator was recently considered in
\cite{Berecotta} in the matrix model including the 3 complex scalar fields
of ${\cal N}=4$ SYM. It would be very interesting to understand the
relation between the manifold supporting the distribution of eigenvalues
in that case and the LLM coordinates.},
leads to $h_n=(n+M)!$ and the eigenvalue distribution,
\begin{eqnarray}
\langle \rho(z) \rangle_N = e^{-|z|^2}|z|^{2M}\sum_{i=0}^{N-1}
\frac{|z|^{2i}}{(i+M)!}\,,
\label{denn}
\end{eqnarray}
which is only distinguishable from (\ref{den}) for finite values of
$N$ and $M$ (see Figure \ref{fig}).

Let us summarize here, a general concentric droplet can be generated
by exciting the ground state with the operator
\begin{eqnarray}
{\cal O}^{(BPS)}_{LLM}\!\!&=&\!\!\prod_{\alpha=1}^M\det(Z-\lambda_\alpha)
%\nonumber
%\\
%\!&=&\!\!\!
=\prod_{\alpha_1=1}^{M_1}\!\!\det(Z-\lambda_{\alpha_1})\dots
\!\!\prod_{\alpha_k=1}^{M_k}\!\!\det(Z-\lambda_{\alpha_k}).
\label{excige}
\end{eqnarray}
where each set $\{\lambda_{\alpha_n}\}$ of $M_n$ holes are
distributed homogeneously in circles of radius $r_n$\footnote{As
discussed in the paragraph above, in the large $N$,$M_1$,... $M_k$
limits there is no difference whether the holes are distributed in
circles or annuli.}. The resulting eigenvalue distribution is a
series of disconnected annuli centered at the origin. It would be
interesting to better understand the relation between the operators
${\cal O}^{(BPS)}_{LLM}$ and the Young Tableaux \cite{Btoy,CJR}. For
the time being, what is relevant to us is that in the large $N$
limit (\ref{excige})  leads to concentric eigenvalue distributions

%%%%%%%%%%%%%%%%%%%%%%%%%%%%%%%%%%%%%%%%%%%%%%%%%%%%%%%%%%%%%%%%%%%%%%%%%%%%%%%%%%%%%%%%%%%%
%%%%%%%%%%%%%%%%%%%%%%%%%%%%%%%%%%%%%%%%%%%%%%%%%%%%%%%%%%%%%%%%%%%%%%%%%%%%%%%%%%%%%%%%%%%%
\section{String excitations in concentric droplets}
\label{string}

In the previous section we have seen how certain BPS operators are
associated to droplet pictures in a plane. These operators are
believed to be dual to some supergravity solutions with exactly the
same quantum numbers, and moreover these solutions are also
characterized by identical droplet pictures. A skeptical reader
might object that only qualitative evidence supports this
identification. In this section we will show that by probing these
BPS operators with a non-BPS factor it is possible to reconstruct
some geometrical and topological information of the dual LLM
geometry.

We will consider single traces probing the BPS operators associated
with the concentric droplets of the previous section and interpret
them as closed string probes of concentric LLM backgrounds. We will
take the probing single traces to be in the so-called $SU(2)$
sub-sector of ${\mathcal{N}}=4$ SYM, that is, in addition to  the
chiral $Z$ fields they also contain complex chiral scalar fields
$Y$. The resulting excited operators are generically non-BPS and
the study of their anomalous dimensions will substantiate their
identification with closed string excitations.

To one-loop order, the dilatation operator in the $SU(2)$ sub-sector
is given by \cite{dil}
\begin{equation}
\label{dila} D = D_0 + \lambda D_1+{\cal O}(\lambda^2)\,,
\end{equation}
here $\lambda = \frac{g^2 N}{8\pi^2}$ is the 't Hooft coupling,
$D_0= {\rm tr}(Z\partial_Z+Y\partial_{Y})$ gives the classical
dimension and the one-loop anomalous dimension operator $D_1$ can be
written in the form,
\begin{equation}
\label{d1} D_1 = N^{-1} {\rm
tr}([Z,Y][\partial_Z,\partial_Y])\,.
\end{equation}
The action of the $D_1$ on the set of operators of the form ${\cal
O}=\mathrm{tr} (ZY...){\cal O}^{(BPS)}_{LLM}$ receives two
contributions, the first comes from the action of $\partial_Z$ (as
well as $\partial_{Y}$) on the trace representing the string. The
second contribution comes from the action of $\partial_Z$ on the
product of determinants ${\cal O}^{(BPS)}_{LLM}$. The action of
$\partial_Z$ on the product of determinants gives a sum over the
$\lambda_{\alpha_n}$, which,  in the continuous limit, can in turn
be approximated by contour integrals in the complex plane. The
result is
\begin{eqnarray}
\label{dzdge}
\!\!\!(\partial_Z)^i_j{\cal O}^{(BPS)}_{LLM}
\!\!\!&=&
 \!\!(\partial_Z)^i_j
\!\!
\prod_{\alpha_1=1}^{M_1}\!\!\det(Z-\lambda_{\alpha_1}\!)...
\!\!\prod_{\alpha_k=1}^{M_k}\!\!\det(Z-\lambda_{\alpha_k}\!)
\nonumber\\
\!\!\!\!\!\!\!\!
&=& (M_1 {(P_1)}^i_j+...+M_k
{(P_k)}^i_j){\cal O}^{(BPS)}_{LLM}\,,
\end{eqnarray}
where
\begin{equation}
P_n=\frac{1}{M_n}\sum_{\alpha=1}^{M_n}\frac{1}{Z-\lambda_\alpha}
\simeq \frac{1}{2 \pi i}\oint_{|\lambda|=r_n}
\frac{d\lambda}{\lambda(Z-\lambda)} \,. \label{pro}
\end{equation}
We will see below that the matrix $P_n$  effectively projects the
string excitations to the droplets lying outside $r_n$.

Let us begin by analyzing the doubly-connected annular domain, which
can be generated, as discussed at the end of previous section, by
exciting  the vacuum with ${\cal O}^{(BPS)}_{LLM}=\det(Z)^M$. In
this case (\ref{dzdge}) reduces to
\begin{equation}
\label{dzdge2} (\partial_Z)^i_j\, \det(Z)^M = M {(Z^{-1})}^i_j\,
\det(Z)^M\,,
\end{equation}
A closed string excitation on the annular LLM geometry is
represented by a single trace operator on top of the product of
determinants. Instead of the usual spin chain labeling
\cite{minahan}, we use a generalization of the bosonic labeling
worked out in \cite{sam,BCV,BCV2,CS,ggas}
\begin{equation}
|n_1,\ldots, n_L\rangle \leftrightarrow {\rm tr}(YZ^{n_1}Y\cdots
YZ^{n_L}) \det(Z)^M\, . \label{stringop}
\end{equation}
Here $L$ is the number of $Y$ impurities in the trace and
represents, for the closed string, an additional angular momentum
along a $S^3$ transverse to the LLM plane. An important difference
for the annular distribution compared to the operators representing
closed strings in $AdS_5\times S^5$ is that the integers $n_i$ in
(\ref{stringop}) can be negative. Negative powers of $Z$ are simply
a shorthand notation for derivatives $\partial_Z$ acting on the
product of determinants. We will show that single trace operators
with a large positive  total occupation number $J=\sum n_l$ can be
pictured as excitations traveling along the exterior edge of the
annulus while those with large negative occupation number can be
seen as the excitations traveling along the interior edge.

When one computes the action of the dilatation operator in the large
$N$ limit at the leading order in $N$, one must take into account
the proper normalization of the operators.  The outcome of this
computation\footnote{Some useful results for deriving the action of
$D_1$ are summarized in the appendix.}  is that the action of $D_1$
over the set $|n_1,\ldots, n_L\rangle$ is closed in the large $N$
limit and can be shown to be represented by the following bosonic
lattice Hamiltonian,
\begin{equation}
H=\lambda\sum_{l=1}^L(a_l-a_{l+1})^\dagger (a_l-a_{l+1}) \,
.\label{wbh}
\end{equation}
The cyclicity of the single trace implies $a_{L+1}=a_1$. Moreover,
$a,a^\dagger$ operators acting at different sites commute. The $a$
and $a^\dagger$ operators are shift operators whose action over
positive and negative occupation orthonormal states $|n\rangle$ is,
\begin{equation}
\label{sh} a|n\rangle = \left\{
\begin{array}{c l}
\sqrt{1+\gamma}\ |n-1\rangle & \quad {\rm if\ } n > 0
\\
\sqrt{\gamma}\ |n-1\rangle & \quad {\rm if\ } n \leq 0
\end{array}
\right.
\end{equation}
where $\gamma=\frac{M}{N}$. Note that (\ref{wbh}) coincides with the
bosonic Hamiltonian that describes closed strings in a $SU(2)$
sub-sector of $AdS_5\times S^5$ \cite{sam}. The difference now being
that the shift operators act over additional negatively occupied
states. Of course, by setting $\gamma=0$ one recovers the shift
operators corresponding to the disc \cite{BCV,BCV2}.

In what follows, it is convenient to define a set of coherent states
of the shift operator $a$ defined in eqn. (\ref{sh}) (such that
$a|z\rangle=z|z\rangle$).  In terms of the Fock \{$|n\rangle$\}
basis they are written as
\begin{equation}
|z\rangle =\sum_{n=-\infty}^{-1} \left(\gamma\right)^{-n/2}
z^n|n\rangle+ \sum_{n=0}^{\infty} \left(1+\gamma\right)^{-n/2}
z^n|n\rangle\,. \label{cohes}
\end{equation}
The normalizability of (\ref{cohes}) compels the complex coordinate
$z$ to be restricted to a specific domain of the complex plane,
\begin{equation}
\langle z|z\rangle =  \sum_{n=1}^{\infty}
\left(\frac{\gamma}{|z|^2}\right)^{n} +
 \sum_{n=0}^{\infty}
\left(\frac{|z|^2}{1+\gamma}\right)^{n} \ . \label{cohen}
\end{equation}
The norm $\langle z|z\rangle$ is finite when the geometric sums are
convergent, i.e. in the domain $\sqrt{\gamma}<|z|<\sqrt{1+\gamma}$.
We will see below that it is possible to identify this domain with
the annular droplet defining the LLM geometry.

Let us remark that the coherent state basis is very useful to
identify the ground state of Hamiltonian (\ref{wbh}). A state
constructed by having the same coherent state in all $L$ sites of
the lattice $|\phi\rangle=|z,\ldots,z\rangle$, is a zero eigenvalue
eigenstate and therefore the ground state of (\ref{wbh}). Being the
total occupation number conserved, this means a fixed total number
of $Z$s in the excitation (\ref{stringop}), it is possible to choose
the Hamiltonian eigenstates to have definite total occupation
number. We call this conserved number $J$. For instance, take the
two-site Hamiltonian
\begin{equation}
H=2\lambda(a_1^\dagger a_1+ a_2^\dagger a_2 -a_2^\dagger
a_1-a_1^\dagger a_2) \, . \label{2sh}
\end{equation}
The ground state with positive total occupation number $J$ can be
obtained as the contour integral $|\phi_0^{(J)}\rangle\propto\oint
dz\ z^{-J-1} |z,z\rangle$. In terms of Fock states it takes the form
\begin{equation}
 |\phi_0^{(J)}\rangle= \sum_{m=1}^\infty
\left(\!\frac{\gamma}{1+\gamma}\!\right)^{\frac{m}2}\!\!\!
(|J+m,-m\rangle + |-m,J+m\rangle) +\sum_{m=0}^J  |J-m,m\rangle\, .
\end{equation}
Similarly, for $-J$ occupation number  one has
\begin{equation}
|\phi_0^{(-J)}\rangle= \sum_{m=1}^\infty
\left(\!\frac{\gamma}{1+\gamma}\!\right)^{\frac{m}2}\!\!\!
(|-J-m,m\rangle + |m,-J-m\rangle) +\sum_{m=0}^J  |m-J,-m\rangle\, .
\end{equation}
The discrete spectrum immediately above the ground state will be
essential for conveying the localization of the near-BPS excitations
at each edge. Moreover, we would like to take a limit in which the
one-loop Hamiltonian description of the anomalous dimension can be
safely extrapolated to strong coupling (a BMN-like limit). There are
two interesting possibilities that allow for explicit comparisons
with string theory computations. The first one is to keep the number
of sites (the number of impurity fields $Y$ in the single trace)
finite while taking the total occupation to infinity $|J|\to\infty$
in such a way that $\lambda/J^2 \ll 1$ \cite{BMN}. An alternative
limit is to take the number of sites $L\to\infty$ in such a way that
$\lambda/L^2 \ll 1$. This second possibility correlates with a
semi-classical (large quantum numbers) string description
\cite{GKP2}.

Let us first look for the spectrum of the two-site Hamiltonian
(\ref{2sh}). In the large $J$ occupation limit, its eigenstates are
BMN operators with two impurities \cite{BMN}. The spectrum is found
by looking for normalizable eigenstates. Take the following ansatz
for a state with total positive occupation $J$
\begin{equation}
|\phi\rangle = \sum_{m=1}^\infty
\left(\!\frac{\gamma}{1+\gamma}\!\right)^{\frac{m}2}\!\!\!
(f_{-m}|J+m,-m\rangle + f_{J+m}|-m,J+m\rangle) +\sum_{m=0}^J f_m
|J-m,m\rangle\ \,, \label{ans}
\end{equation}
Requiring (\ref{ans}) to be an eigenstate of (\ref{2sh}) leads to a
recurrent second order equation for the $f_n$ coefficients. One of
the two arbitrary constants of a given solution amounts merely to a
normalization. The second one has to be adjusted to get a
normalizable state. We fix it by imposing the vanishing of
$\left(\frac{\gamma}{1+\gamma}\!\right)^{n/2}f_{J+n}$ as $n$ goes to
infinity. The vanishing  of
$\left(\frac{\gamma}{1+\gamma}\!\right)^{n/2}f_{-n}$ as $n$ goes to
infinity requires to fine-tune the eigenvalue. The Hamiltonian
eigenvalues of normalizable eigenstates must then
satisfy\footnote{The energy appearing in (\ref{con}) is given in
units of $2\lambda(1+\gamma)$.},
\begin{eqnarray}
0&=&(2-\sqrt{E^2-4E}-E)^J\left[-E^2(1+\gamma)^2
+ E(3+7\gamma +4\gamma^2)\right.\nn \\
&&+\left.(1+(1+\gamma)\sqrt{E^2-4E})
(-1+\sqrt{1+E^2 (1+\gamma)^2-2E(1+3\gamma +2\gamma^2)})\right]\nn\\
&&+ (2+\sqrt{E^2-4E}-E)^J\left[ E^2(1+\gamma)^2
- E(3+7\gamma +4\gamma^2)\right.\nn \\
&&+\left.(1-(1+\gamma)\sqrt{E^2-4E}) (1-\sqrt{1+E^2
(1+\gamma)^2-2E(1+3\gamma +2\gamma^2)})\right]\,,
\label{con}
\end{eqnarray}
Equation (\ref{con}) is transcendental, but a solution for $E$ can
be found assuming an expansion in powers of $1/J$. The result is
\begin{equation}
E^{(+)}_n=(1+\gamma)\frac{8\pi^2 n^2 \lambda}{J^2}\left(
  1-\frac{2+4\gamma}{J}+{\cal
  O}\left(\frac{1}{J^2}\right)\right)\, ,
\label{bmn1}
\end{equation}
where $n$ is an integer. A similar analysis shows that for large
negative occupation number $-J$, the energy for the first excited
states is
\begin{equation}
E^{(-)}_n=\gamma\frac{8\pi^2 n^2 \lambda}{J^2}\left(
  1-\frac{2+4\gamma}{J}+{\cal O}\left(\frac{1}{J^2}\right)\right)\, .
\label{bmn2}
\end{equation}

This is our first important result. The string spectrum on the
plane-wave spacetime obtained, as a Penrose limit, when zooming
around null geodesics sitting on the exterior and interior edges of
the LLM annular droplet are actually different (see (\ref{ema}) and
(\ref{eme}) in the appendix). The matching of the leading terms of
(\ref{bmn1}) and (\ref{bmn2}) with (\ref{ema}) and (\ref{eme}) is
perfect. This is a clear evidence that a single trace excitation
with a large positive number of $Z$'s and a finite number of $Y$'s
can be regarded as a  string excitation traveling along the exterior
edge; whereas a single trace operator with a large negative number
of $Z$'s and finite $Y$'s  can be regarded as a string excitation
traveling along the interior edge. It also gives confidence for the
correctness of the BPS operators (\ref{excige}) claimed in section
\ref{bps} to describe the LLM geometries.

The fact that the coherent state complex coordinate domain is an
annulus indicates that we can  attribute a geometrical meaning to
it: the domain of the complex coordinate associated to the coherent
states coincides with the support of droplet defined on the LLM
($y=0$) plane \cite{sam}.

To give support to this last claim we will now extrapolate the
one-loop analysis to strong coupling regime performing the second
BMN-like limit mentioned above. We will take the number of sites $L$
to infinite keeping $\lambda/L^2$ fixed and small. This limit is
known, in the context of integrable spin-chains, as the
``Thermodynamic limit'' \cite{kru,ss}.

On the string side, this second limit corresponds to consider
semi-classical string solutions in the annulus background. Consider
in particular a folded string stretching between $R_1$ and $R_2$ in
the LLM plane and carrying a large angular momentum  $L$ in the
transverse $S^3$ (see the appendix). This angular momentum $L$
should be identified with the $U(1)_Y$ $R$-charge carried by the $Y$
chiral fields. The one-loop energy $E$, in asymptotic global $AdS$
coordinates, of this folded string solutions can be expressed, in
the large $L$ limit, as (see (\ref{esfoldex}))
\begin{equation}
E - J- L \simeq
\frac{4\lambda}{L}(\sqrt{1+\gamma}-\sqrt{\gamma})^2\,.
\label{efold}
\end{equation}
We will now reproduce this relation from a semiclassical gauge
theory computation. Take a lattice coherent state $|z_1\ldots
z_L\rangle$, where the $z_l$ are arranged as
\begin{equation}
z_l = \left\{
\begin{array}{ll}
(\sqrt{1+\gamma}-\sqrt{\gamma})\frac{2l}{L} + \sqrt{\gamma} & {\rm
if\ }  l \leq L/2
\\
(\sqrt{\gamma}-\sqrt{1+\gamma})\frac{2l}{L} +2\sqrt{1+\gamma} -
\sqrt{\gamma} & {\rm if\ } l > L/2
\end{array}
\right. \label{zzz}
\end{equation}
In the continuum limit this election mimics the parametrization of
the folded string solution. The expectation value of the Hamiltonian
on this state exactly coincides with the leading term in the
BMN-like expansion (\ref{efold})
\begin{equation}
\langle z_1... z_L |H|z_1... z_L\rangle =
\frac{4\lambda}{L} (\sqrt{1+\gamma}-\sqrt{\gamma})^2\,.
\label{hfold}
\end{equation}
This is just a particular case of a  general and more notable
matching: the semi-classical sigma-model action corresponding to the
one-loop Hamiltonian describing the system in the large $L$
(continuum) limit coincides with the Polyakov action for a string
propagating in corresponding LLM geometry in the gauge where the
angular momentum $L$ on the $S^3$ is homogeneously distributed along
the string\footnote{The choice of gauge is motivated by the bosonic
labeling (\ref{stringop}) which distributes uniformly the $Y$
fields.} and when a fast-string limit is taken. The semi-classical
sigma model action  is given by
\begin{equation}
S=\int\! dt\left(i \langle z_1... z_L|\frac{d}{dt}|z_1...
z_L\rangle -\langle z_1... z_L|H|z_1... z_L\rangle \right).
\label{coac}
\end{equation}
In the large $L$ limit, the sums from 1 to $L$ appearing in
(\ref{coac}) can be approximated by integrals. The result is
\begin{equation}
S=L \int\!dt \int_0^1\!d\sigma\left(\frac{i}{2} V \dot{ \bar z}
-\frac{i}{2} \bar V\dot{ z} -\frac{\lambda}{L^2}|z'|^2\right)\,,
\label{coac2}
\end{equation}
where
\begin{eqnarray}
V(z,\bar z) &\!\!= \!\!&\partial_{\bar z}\log(\langle
z|z\rangle)\nn
\\
&\!\!= \!\!& \frac{z}{\hat R_2^2-|z|^2} + \frac{\hat R_1^2}{{\bar
z}(\hat R_1^2-|z|^2)}\, , \label{vf}
\end{eqnarray}
with $\hat R_1=\sqrt{\gamma}$ and $\hat R_2=\sqrt{1+\gamma}$. This
is our second important result for the annulus: The action
(\ref{coac2}) coincides with the  fast-string limit of the Polyakov
action, when written in a gauge that homogeneously distributes along
the string the angular momentum on the transverse $S^3$
(cf.(\ref{rsllm})). Moreover, the expression for the function
$V(z,\bar z)$ in (\ref{coac2}) reproduces exactly the LLM function
$V(z,\bar z)$  for the annulus droplet (cf.(\ref{ves})). As a
conclusion, by probing the operator $\det(Z)^M$ we were able to
recreate some information about the metric of the LLM annular
solution (the one-form $V_a$ restricted to the LLM plane $y=0$).

~

\begin{figure}%[htb]
\begin{center}
\epsfxsize=2.5in\leavevmode\epsfbox{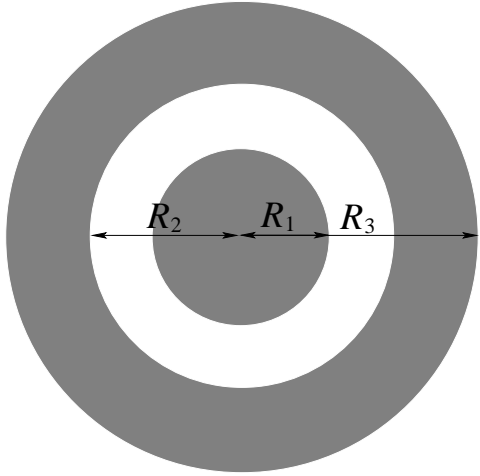}
\end{center}
\caption{Distribution picture of $N_1$ eigenvalues in a disc  and
$N_2$ in the annulus.} \label{fig2}
\end{figure}

In the following we will study the generalization corresponding to
distributing $M$ holes in a circle  centered around the origin. As
discussed at the end of section \ref{bps} this leads to a droplet
consisting of a black disc and an annulus (see Figure \ref{fig2}).
The eigenvalues therefore get distributed in two disconnected
droplets: a disc of radius $R_1=\sqrt{N_1}$ and an annular droplet
of radii $R_2=\sqrt{N_1+M}$ and $R_3=\sqrt{N_1+N_2+M}$. Now, not
only should one identify different near-BPS excitations associated
to the three edges, but also be able to determine if a given
semi-classical non-BPS excitations is localized either in the disc
or in the annulus.

As we did for our first example, we analyze the action of
$\partial_Z$  on the product of determinants
\begin{equation}
\label{dzdge3}
(\partial_Z)^i_j
\,{\cal O}^{(BPS)}_{LLM}=(\partial_Z)^i_j\,
\prod_{\alpha=1}^M\det(Z-\lambda_\alpha)=
M {(P)}^i_j \,{\cal O}^{(BPS)}_{LLM}\,.
\end{equation}
The resulting  $P$  matrix can be expressed in terms of the
eigenvalues of the exterior droplet. To see this, it is convenient
to use a matrix $\Lambda$ that diagonalizes $Z$ \footnote{Since
$Z$ is complex rather than normal, $\Lambda$ is in general
non-unitary.}
\begin{eqnarray}
(P)^i_j \!&\simeq&\! \frac{1}{2 \pi i}\sum_{k=1}^N
(\Lambda)^i_k(\Lambda^{-1})^k_j
\oint_{|\lambda|=r_n}\!\frac{d\lambda}{\lambda(z_k-\lambda)}\nn
\\
\! &=& \! \sum_{I/ |z_I|>
r}\frac{(\Lambda)^i_I(\Lambda^{-1})^I_j}{z_I}\,.
\label{pro2}
\end{eqnarray}
In particular, note that $PZ$ is a projector, satisfying
$(PZ)^{2}=PZ$. Let us remark in passing that the action of $D_0$ on
the product of determinants in (\ref{dzdge3}) gives $M\,{\rm
tr}(PZ)$ times the product of determinants and note that $M{\rm
tr}(PZ)$ gives just a numerical factor $MN_2$. This is a reassuring
result since one expects the BPS operator corresponding to a
concentric eigenvalue distribution to have a definite number of $Z$
fields.

The presence of the operator $P$ suggests to write a general closed
string excitation as
\begin{equation}
|\phi_1,\ldots, \phi_L\rangle \leftrightarrow {\rm
tr}(Y\phi_1(Z)Y\cdots Y\phi_L(Z))
\prod_{\alpha=1}^M\det(Z-\lambda_\alpha)\,, \label{stringop2}
\end{equation}
where the possibilities for $\phi_l(Z)$ are
\begin{equation}
\phi_l(Z)= \left\{
\begin{array}{ll}
P^n & {\rm for\ } n\geq 1
\\
Z_P^n\equiv Z^{n+1}P & {\rm for\ } n\geq 0
\\
Z_Q^n\equiv Z^n - Z^{n+1}P & {\rm for\ } n\geq 0
\end{array}
\right. .\label{basis}
\end{equation}
Here the products $P^n$ can be thought of as negative powers of $Z$
with eigenvalues only belonging to (or projected onto) the annular
droplet, while $Z_P^n$ and $Z_Q^n$ are thought of as positive powers
of $Z$ projected onto the annulus and the disc respectively. The
conclusion is that the appearance of the $P$ operator opens the
possibility of occupying each site in the lattice with three
different types of bosons.

The next step is to study the action of the one-loop dilatation
operator (\ref{d1}) over the set of operators (\ref{stringop2}).
However, the basis (\ref{basis}) is not the most appropriate one
since states occupied by $Z_P^n$ and $Z_Q^n$ are not orthogonal.
This may seem odd a first sight, since $Z_P^n$ is projected with
$PZ$ and $Z_Q^n$ by its complement. Their interior product is
\begin{equation}
\langle Z_P^n | Z_Q^n \rangle \simeq \langle {\rm
tr}(\overline{Z_P^n} Z_Q^n) \rangle \, .
\end{equation}
While $Z_P^n Z_Q^n$ is clearly zero, $\overline{Z_P^n} Z_Q^n$ is
not. This is due to the matrix $Z$  being complex and $\Lambda$
being non-unitary in general. The set of orthogonal states we will
consider is
$\{|P^n\rangle,|Z_P^0\rangle,|n\rangle,|0\rangle,|Z_Q^n\rangle  \}$
where
\begin{eqnarray}
&&  |0\rangle \equiv (N_1+M) |Z_Q^0\rangle + N_1 |Z_P^0\rangle
\, ,
\\
&&  |n\rangle \equiv (N_2+M) |Z_P^n\rangle + N_2 |Z_Q^n\rangle \quad
{\rm for\ } n \geq 1\, , \label{ortobasis}
\end{eqnarray}
The leading contribution to the action of $D_1$, when acting on the
set discussed in the previous paragraph is again represented by a
Hamiltonian of the form (\ref{wbh}),
\begin{equation}
H=\lambda\sum_{l=1}^L(a_l-a_{l+1})^\dagger (a_l-a_{l+1}) \, ,
\label{wbh2}
\end{equation}
where the shift operators now act differently depending on the type
of boson occupancy\footnote{In all formulae
(\ref{sh21})-(\ref{sh27}) the states should be understood as
orthonormal.},
\begin{eqnarray}
\label{sh21} a|P^n\rangle \!&=&\! \sqrt{\nu_1+\gamma}\
|P^{n+1}\rangle  \quad \quad {\rm if\ } n \geq 1\, ,
\\
\label{sh22} a |Z_P^0 \rangle \!&=&\! \sqrt{\nu_1+\gamma}
|P\rangle\, ,
\\
\label{sh23} a|0\rangle \!&=&\! 0\, ,
\\
\label{sh24} a|Z_Q\rangle \!&=&\!
\sqrt{\frac{\nu_1\gamma(1+\gamma)}{(\nu_1+\gamma)(\nu_2+\gamma)}}
|0\rangle - \sqrt{\frac{\nu_1^2
\nu_2}{(\nu_1+\gamma)(\nu_2+\gamma)}} |Z_P^0\rangle\, ,
\\
\label{sh25} a|1\rangle \!&=&\!
\sqrt{\frac{\nu_1\nu_2(1+\gamma)}{(\nu_1+\gamma)(\nu_2+\gamma)}}
|0\rangle +
\sqrt{\frac{\gamma(1+\gamma)^2}{(\nu_1+\gamma)(\nu_2+\gamma)}}
|Z_P^0\rangle\, ,
\\
\label{sh26} a|Z_Q^n\rangle \!&=&\!  \sqrt{\nu_1}\ |Z_Q^{n-1}\rangle
\quad \quad \quad \quad{\rm if\ } n \geq 1\, ,
\\
\label{sh27} a|n\rangle \!&=&\!  \sqrt{1+\gamma}\ |{n-1}\rangle
\quad \quad \ {\rm if\ } n \geq 1\, ,
\end{eqnarray}
here $\gamma=\frac{M}{N}$, $\nu_1=\frac{N_1}{N}$ and
$\nu_2=\frac{N_2}{N}$. As before, it is  useful, for the
semi-classical description, to define coherent states of the shift
operator $a$. However, there is now no unique way of doing so. It is
now possible to define coherent states involving no $|P^n\rangle$,
\begin{eqnarray}
\!\!|z\rangle_{\rm I} \!&=&\! |0\rangle+
\sqrt{\frac{\gamma(1+\gamma)}{(\nu_1+\gamma)(\nu_2+\gamma)}}
\sum_{n=1}^{\infty}\left(\frac{z}{\sqrt{\nu_1}}\right)^n\! 
|Z_Q^n\rangle \nonumber\\ &&
+\sqrt{\frac{\nu_1\nu_2}{(\nu_1+\gamma)(\nu_2+\gamma)}}
\sum_{n=1}^{\infty}
\left(\frac{z}{\sqrt{1+\gamma}}\right)^n\! |n\rangle. \label{cohesI}
\end{eqnarray}
Alternatively, it is also possible to define coherent states
involving no $|Z_Q^n\rangle$,
\begin{eqnarray}
&& |z\rangle_{\rm II} = |0\rangle+
\sqrt{\frac{\gamma(1+\gamma)}{\nu_1\nu_2}}
 |Z_P^0\rangle+
\sqrt{\frac{\gamma(1+\gamma)}{\nu_1\nu_2}}
\sum_{n=1}^{\infty}\left(\frac{z}{\sqrt{\nu_1+\gamma}}\right)^{-n}
|P^n\rangle \nn
\\
&&\qquad\qquad +
\sqrt{\frac{(\nu_1+\gamma)(\nu_2+\gamma)}{\nu_1\nu_2}}
\sum_{n=1}^{\infty}
\left(\frac{z}{\sqrt{1+\gamma}}\right)^n|n\rangle\ . \label{cohesII}
\end{eqnarray}
Both (\ref{cohesI}) and (\ref{cohesII}) satisfy
$a|z\rangle=z|z\rangle$. Moreover (\ref{cohesI}) and(\ref{cohesII})
are normalizable in different complimentary domains,
\begin{equation}
 _{\rm I}\langle z|z\rangle_{\rm I} = 1+
\frac{\nu_1\nu_2}{(\nu_1+\gamma)(\nu_2+\gamma)}
\sum_{n=1}^{\infty}\frac{|z|^{2n}}{\nu_1^{n}} +
\frac{\gamma(1+\gamma)}{(\nu_1+\gamma)(\nu_2+\gamma)}
\sum_{n=1}^{\infty}\frac{|z|^{2n}}{(1+\gamma)^{n}}, \label{cohenI}
\end{equation}
which is finite for $|z|<\nu_1$, the disc region, and
\begin{equation}
 _{\rm II}\langle z|z\rangle_{\rm II} =
1+ \frac{\gamma(1+\gamma)}{\nu_1\nu_2}
\sum_{n=0}^{\infty}\frac{(\nu_1+\gamma)^{n}}{|z|^{2n}} +
\frac{(\nu_1+\gamma)(\nu_2+\gamma)}{\nu_1\nu_2}
\sum_{n=1}^{\infty}\frac{|z|^{2n}}{(1+\gamma)^{n}}, \label{cohenII}
\end{equation}
which is finite for $\nu_1+\gamma<|z|<1+\gamma$, the annulus region.

When using $|z\rangle_{\rm I}$ to compute the semi-classical action
as done in (\ref{coac})-(\ref{vf}), one finds (\ref{coac2}) again.
The only difference is in the function $V(z,\bar z)$, which in the
present case takes the form
\begin{eqnarray}
V(z,\bar z) &\!\!= \!\!&\partial_{\bar z}\log(_{\rm I}\langle
z|z\rangle_{\rm I})\nn
\\
&\!\!= \!\!& \frac{z}{\hat R_3^2-|z|^2} - \frac{z}{\hat R_2^2-|z|^2}
+ \frac{z}{\hat R_1^2-|z|^2}\, , \label{v1}
\end{eqnarray}
where ${\hat R}_1 = \sqrt{\nu_1}$, ${\hat R}_2 =
\sqrt{\nu_1+\gamma}$ and ${\hat R}_3 = \sqrt{1+\gamma}$. This is
exactly the function $V(z,\bar z)$ of a concentric droplet with
three edges, for $|z|<{\hat R}_1$ (see (\ref{ves}))

Alternatively, when using $|z\rangle_{\rm II}$, a similar result is
obtained for the semiclassical action, the function $V(z,\bar z)$ in
the  (\ref{coac2}) now being
\begin{eqnarray}
V(z,\bar z) &\!\!= \!\!&\partial_{\bar z}\log(_{\rm II}\langle
z|z\rangle_{\rm II})\nn
\\
&\!\!= \!\!& \frac{z}{{\hat R}_3^2-|z|^2} + \frac{z}{{\hat
R}_2^2-|z|^2} - \frac{z}{{\hat R}_1^2-|z|^2}\, . \label{v2}
\end{eqnarray}
This is $V(z,\bar z)$ as in (\ref{ves}) for ${\hat R}_2<|z|<{\hat
R}_3$.

In conclusion, the two different coherent states $|z\rangle_{\rm I}$
and $|z\rangle_{\rm II}$ allow for two different large $L$
semiclassical limits. The first one valid for the disc $|z|<{\hat
R}_1$ of the complex plane, while the second is valid for the
annulus ${\hat R}_2<|z|<{\hat R}_3$. The two semi-classical actions
so obtained for the bosonic lattice again coincide with the Polyakov
action in the fast-string limit. Using $|z\rangle_{\rm I}$ leads to
semiclassical strings stretching inside the disc and using
$|z\rangle_{\rm II}$  to semiclassical strings within the black
annulus.

%%%%%%%%%%%%%%%%%%%%%%%%%%%%%%%%%%%%%%%%%%%%%%%%%%%%%%%%%%%%%%%%%%%%%%%%%%%%%%%%%%%%%%%%%%%%
%%%%%%%%%%%%%%%%%%%%%%%%%%%%%%%%%%%%%%%%%%%%%%%%%%%%%%%%%%%%%%%%%%%%%%%%%%%%%%%%%%%%%%%%%%%%
\section{Discussion}
\label{discu}

In this paper, we have derived in the large $N$ limit, the action of
the one-loop dilatation operator on the set of gauge theory
operators representing closed strings probing axially symmetric
bubbling geometries whose corresponding droplets can have multiple
edges and non-connected constituents. BMN-like limits were taken to
extrapolate  reliably the one-loop computations to strong 't Hooft
coupling and account for both geometrical and topological features
of the bubbling geometries.

Concerning the topology of the SUGRA solutions, let us first remark
that the gauge theory description was able to account for different
sorts of BPS and near-BPS excitations that appear for multiple-edged
droplets. The operators we found were in obvious correspondence with
the dual string excitations traveling along null or almost
null-geodesics (corresponding to the trajectories of BMN
excitations) present on the droplet edges.

Motivated by \cite{sam},  a large number of impurity fields $Y$ was
inserted in the single trace operator corresponding to the closed
string probe. In doing so, the gauge theory description reproduced
the notion of LLM coordinates and was also able to reproduce some of the
functions characterizing the bubbling metric. The novel
characteristic for droplets with non-connected constituents was that
different but complimentary semi-classical descriptions of the
one-loop Hamiltonian were allowed. They precisely coincided with the
actions of the possible semi-classical string solutions extending
in each of the different disconnected components of the droplet with
a fast-string limit imposed.

We specifically considered two bubbling configurations: the annulus,
which is connected but has two edges, and the superposition of a
disc and an annulus which is the simplest non-connected droplet.

Relying on the relation of $\frac12$-BPS states of ${\cal N}=4$ SYM
to matrix models \cite{Btoy,CJR}, we gave a prescription for the
dual $\frac12$-BPS operators to any arbitrary concentric droplet.

It has been previously seen that the action of the one-loop
dilatation operator on the $SU(2)$ sub-sector operator can be
reformulated in terms of the Hamiltonian of a bosonic lattice
\cite{BCV,BCV2}. In the first (annulus) example, we considered a
single trace operator on top of $\det(Z)^M$. The Hamiltonian we
obtained acts over a lattice whose sites can be occupied either by a
positive or negative number of bosons. We explicitly solved the
Hamiltonian spectrum for the case of large total boson occupation
number, distinguishing the two possible cases of positive and
negative occupations. In both cases the result was a BMN-like
spectrum and the spectrum for positive (negative) occupation
reproduced exactly the one obtained by quantizing a closed string
traveling in the almost null-geodesic associated with the exterior
(interior) edge of the annulus.

We remark that the axial symmetry of the droplets was essential for
successfully computing the spectrum of the Hamiltonian. The axial
symmetry entails a conserved total number of bosons. These
eigenstates with definite number of bosons turn out to be solutions
of a simple second order recurrent equation.

We also showed that the semi-classical sigma-model action for a
lattice with a large number of sites coincides with a fast-string
limit of the Polyakov action, when parameterized in such a way such
that the angular momentum dual to the $R$-charge of the $Y$ fields
is uniformly distributed along the string. This is in accordance
with the bosonic lattice labeling of the probing single trace, since
it uniformly accounts the $Y$ fields. In doing so, we were able to
re-create the one-form $V_a$ of the LLM metric when restricted to the
droplet plane $y=0$.

In the second (disc+annulus) example, the resulting Hamiltonian
could again be expressed as acting on a bosonic lattice. In this
case, the sites could be occupied by three different kinds of
bosons. This was the gauge theory realization of the droplet having
three edges. These corresponded to powers of the chiral complex $Z$
field projected either into the disc or into the annulus. It is
significant to remark that the expression of the Hamiltonian is the
same for all concentric cases, the difference being in the shift
operators used. The action of the shift operators encodes the number
of edges in the droplet and their radial distances to the origin.
Moreover, for non-connected droplets, the set of coherent states
of the shiftoperator is not unique. In the disc+annulus example,
there were two possible  types of coherent states $|z\rangle_{\rm I}$
in the domain $|z|<\hat R_1$, and
$|z\rangle_{\rm II}$ in the domain $\hat R_2<|z|<\hat R_3$. The
semi-classical descriptions derived from them coincided with the
description of classical closed strings extended inside the disc or
within the annulus of the droplet respectively.

Finally, it is interesting to mention that the isometry group $R\times
SO(4)\times SO(4)$ of general LLM geometries is in fact the full
bosonic component of $PSU(2|2)\times PSU(2|2)\ltimes R$. In the spin
chain corresponding to $AdS_{5}\times S^{5}$, this is the residual
symmetry of the ferromagnetic ground state. Short representations of
the symmetry group extended to $PSU(2|2)\times PSU(2|2)\ltimes
R^{3}$ give exact dispersion relations for elementary magnon
excitations in the spin chain \cite{Beisert:2005tm} as well as their
bound states \cite{CDO2}. Open string solutions stretched between
two points on the disc edge are interpreted as giant magnons
\cite{HM} and magnon bound states \cite{Dorey}, where the length of
the string is related to the momentum carried by the magnon
excitation. One can naturally ask if excitations with  similar exact
dispersion relation can exist in a general LLM background. Such
solutions should again be an open string with ends at the edges of a
general droplet.  In this regard, we have found that there are
semiclassical folded strings solutions in general LLM droplets,
whose embedding is a straight line in the droplet with both folding
points localized at edges. We could interpret our folded strings as
a pair of  generalized magnon bound states \cite{Dorey}. A crucial
element in completing this argument is to construct scattering
matrix between the generalization of elementary magnons, hence
identifying   the required pole for the bound states. We hope to return to
this issues in near future.

\section*{Acknowledgements}
The authors are grateful to D.Berenstein, N.Dorey and S.V\'azquez
for useful discussions and comments. H.Y.C. would like thank H. Lin
for the stimulating discussions at the early stage of this project,
he would also like to thank Physics Department, National Taiwan
University for the hospitality during the final stage of the
preparation. D.H.C. work is supported by PPARC grant ref. PP/D507366/1.
G.A.S. would like to thank DAMTP for warm hospitality
at the early stages of this work and acknowledges support from
CONICET, PIP 6160.

%%%%%%%%%%%%%%%%%%%%%%%%%%%%%%%%%%%%%%%%%%%%%%%%%%%%%%%%%%%%%%%%%%%%%%%%%%%%%%%%%%%%%%%%%%%%
%%%%%%%%%%%%%%%%%%%%%%%%%%%%%%%%%%%%%%%%%%%%%%%%%%%%%%%%%%%%%%%%%%%%%%%%%%%%%%%%%%%%%%%%%%%%
\section*{Appendix}

{\bf LLM geometries}

In \cite{LLM} Lin, Lunin and Maldacena worked out the regular 1/2
BPS solutions of type IIB supergravity with isometry group $R\times
SO(4)\times SO(4)$. The geometric content of the solutions is given
by the following metric.
\begin{equation}
ds^2= -h^{-2}(dt + V_a dx_a)^2 + h^2(dy^2+dx_a^2)+ y e^G
d\Omega_3^2+y e^{-G} d{\tilde \Omega}_3^2\,,
\end{equation}
where
\begin{equation}
h^{-2}= 2 y \cosh G\, ,\quad  \tanh G = 2 \rho(x_1,x_2,y) +1\,.
\end{equation}
In general, the functions $V_a(x_1,x_2,y)$ and $\rho(x_1,x_2,y)$ are
obtained by solving a linear differential equation with boundary
conditions $\rho(x_1,x_2,0)$ specified on the  two dimensional $y=0$
(droplet) plane. A smooth non-singular geometry requires
$\rho(x_1,x_2,0)$ to take the values 0 or $-1$.

A concentric droplet with $K$ edges can be analytically solved, the
functions $V_a$ and $\rho$ adopt the form
\begin{eqnarray}
&&\!\!\!\!\! \rho(x_1,x_2,y) =  \sum_{k=1}^K
\frac{(-1)^{K-k}}{2}\!\left(
\frac{r^2+y^2-R_k^2}{\sqrt{(r^2+y^2+R_k^2)^2- 4 r^2
R_k^2}}-1\right),
\\
&&\!\!\!\!\!V_\phi(x_1,x_2,y) = \sum_{k=1}^K
\frac{(-1)^{K-k+1}}{2}\!\left(
\frac{r^2+y^2+R_k^2}{\sqrt{(r^2+y^2+R_k^2)^2- 4 r^2 R_k^2}}-1\right)
. \label{v}
\end{eqnarray}
When the droplet is just the disc, the geometry is $AdS_5\times S^5$
and any compact droplet has $AdS_5\times S^5$ asymptotics. Notice that 
with a series of concentric annuli a concentric droplet with infinite 
black area can be  constructed. In that case the resulting geometry 
possesses a different causal boundary \cite{mosaffa}.

~

\noindent{\bf Penrose limits and string spectra}

The Penrose limits of LLM geometries corresponding to concentric
droplets were computed in \cite{Ebrahim,pp}, where the case of the annular
droplets becoming thinner as the limit was taken.
However, we are interested in the case where all the radii of the
concentric domains are taken large while keeping their ratios fixed. The
resulting plane wave geometries turn out to be the maximally
supersymmetric plane-wane \cite{blau}. Nevertheless, it is important
to keep track of the change of coordinates used in zooming a
particular null geodesic of the LLM geometry in order to make a
comparison among the string and gauge theory sides.

We consider the following change of coordinates in a generic LLM
concentric droplet,
\begin{eqnarray}
\label{cc1}
&& t=u\, , \qquad\qquad \qquad \phi = \pm \frac{v}{R_i}\,,
\\
&& r = R_i +\frac{1}{2}(r_1^2-r_2^2)\,,\quad y = r_1 r_2\,,
\label{cc2}
\end{eqnarray}
where $R_i$ is the radius of any of the edges. The positive sign in
(\ref{cc1}) is chosen for the case where the edge separates an
interior white region from an exterior black one. The negative sign
 is chosen for the opposite case. In both cases, taking all the
$R_i$ to infinity, the resulting metric is the usual pp-wave
\begin{equation}
ds^2=-2 du dv -(r_1^2+r_2^2)du^2 + d\vec{r}_1^2 + d\vec{r}_2^2\, .
\label{ppw}
\end{equation}
At this point we can just borrow the spectrum of $H_{lc}$
corresponding to a closed string in the background (\ref{ppw})
\cite{met},
\begin{equation}
H_{lc}=-p_u=\sum_n N_n \sqrt{1+ \frac{ n^2}{(\alpha' p_v)^2}}
\,. \label{spec}
\end{equation}
We can relate the charges $p_u$ and $p_v$ to the charges $E$ and
$\hat J$ defined in coordinates that asymptotically tend to global
$AdS_5\times S^5$ coordinates. Defining $\tilde t =t= u$ and $\tilde
\phi = t+\phi= u \pm \frac{v}{R_i}$, we use
\begin{eqnarray}
&&-p_u=i\partial_u = i(\partial_{\tilde t}+\partial_{\tilde
\phi}) =E-\hat J\,,
\\
&&-p_v=i\partial_v = \pm \frac{i}{R_i}\partial_{\tilde \phi}
=\mp\frac{\hat J}{R_i} \,,
\end{eqnarray}
to cast the spectrum (\ref{spec}) in the form,
\begin{equation}
E-\hat J=\sum_n N_n \sqrt{1+ \frac{ n^2 R_i^2}{{\alpha'}^2 J^2}}
=\sum_n N_n \sqrt{1+\frac{\pi R_i^2}{{\rm Area}(D)} \frac{8\pi^2 n^2
\lambda}{J^2}} \,,
\end{equation}
where we have also used that Area$(D)=4 \pi^2 g_s N {\alpha'}^2$.

Let us consider for instance the annular droplet. Choosing $R_i=R_2$
in (\ref{cc1})-(\ref{cc2}), the Penrose limit zooms around the
exterior edge. We use $\frac{\pi R_2^2}{{\rm Area}(D)}= 1+\gamma$
and we let the angular momentum $\hat J$ take the value $J$.
Expanding for $\frac{\lambda}{J^2}\ll 1$, we obtain
\begin{equation}
E \simeq J + 2 +(1+\gamma) \frac{8\pi^2 n^2 \lambda}{J^2 }\,.
\label{ema}
\end{equation}
Similarly, the election $R_i=R_1$ zooms around the interior edge. In
this case, $\frac{\pi R_1^2}{{\rm Area}(D)}=\gamma$ and we take the
angular momentum $\hat J$ to be $-J$. Once again, expanding for
$\frac{\lambda}{J^2}\ll 1$, we obtain
\begin{equation}
E \simeq -J + 2 +\gamma \frac{8\pi^3 n^2 \lambda}{J^2 }\,,
\label{eme}
\end{equation}
where $\gamma$ is, as defined before, the ratio between area of the
interior white disc and the area of the black annulus.

~

\noindent{\bf Semi-classical strings}

Semiclassical strings in concentric LLM backgrounds have been
previously studied in \cite{Filev}, \cite{Ebrahim}, \cite{Alishahiha}. There, cases
where the string was extended along $y$ were considered. We are
interested instead in strings propagating in the $y=0$ section. More
precisely, we will consider strings extended in any of the black
regions of the LLM plane and also spinning along an angle $\eta$ of
the transverse $S^3$. We will compare the description of these
strings with large $J_\eta$ with a semiclassical description for the
one-loop Hamiltonian with $L$ taken large. The index parameterizing
the lattice hamiltonian counts uniformly the $Y$ fields of the
single trace probing the BPS background operator. Then, a meaningful
comparison with the string theory description can be done only
fixing the parametrization of the worldsheet in a way such that
$J_\eta$ is uniformly distributed along the string. This particular
gauge fixing \cite{Aru} in the Polyakov action for $AdS_5\times S^5$
written in LLM coordinates was originally done in \cite{BCV} and
more recently for a generic LLM geometry in \cite{sam}, so we only
quote the main result here. It is interesting to distinguish two
possible situations. Firstly, some remarkable simplifications take
place when restricting to static configurations in LLM coordinates,
i.e. when $t=\tau$ and $r$ and $\phi$ only depending on $\sigma$. In
that case, the resulting expression for the action is,
\begin{equation}
\label{staticllm} S = -\frac{L}{2\pi} \int d\tau \int_0^{2\pi}
d\sigma \sqrt{1+\frac{1}{{\alpha'}^2 L^2}({r'}^2 + r^2
{\phi'}^2)}\,,
\end{equation}
Remarkably, all dependence on the LLM function $V_{\phi}(r,\phi)$
disappears. This holds even for the generic (non-axially symmetric)
case. Thus, static solutions correspond to straight lines in the LLM
droplet. Consider a folded string solution stretching between $R_i$
and $R_{i+1}$, so that its turning points move along light-like
trajectories. This solution has the dispersion relation:
\begin{equation}
\label{esfold}
E -J =
L\sqrt{1+\frac{(R_{i+1}-R_{i})^2}{\pi^2{\alpha'}^2 L^2} }=
\sqrt{L^{2}+\frac{(R_{i+1}-R_{i})^{2}}{\pi^{2}{\alpha'}^{2}}}\,.
\end{equation}
The dispersion relation for the folded string in fact closely
resembles the one for magnon bound states \cite{Dorey}\footnote{The
reader should note that in \cite{Dorey}, $\lambda=g_{s}N$, whereas
here we define $\lambda=\frac{g_{s}N}{8\pi^{2}}$}. This can be made
more transparent by introducing the normalized distance.
\begin{equation}
(\hat{R}_{i+1}-\hat{R_{i}})^{2}=\frac{\pi(R_{i+1}-R_{i})^{2}}{{\rm{Area}}(D)}\,,
\label{nord}
\end{equation}
where we have introduced the droplet area Area$(D)=4\pi^2g_s
N{\alpha'}^2 = 8\pi^3\lambda{\alpha'}^2$. Using (\ref{nord}) in
(\ref{esfold}) we have
\begin{equation}
E-J=\sqrt{L^{2}+8\lambda(\hat{R}_{i+1}-\hat{R}_{i})^{2}}\,.
\end{equation}
The folded string being a closed string solution should be thought
of as a combination of two magnon bound states of equal charges but
of opposite quasi-momentum. The magnon quasi-momentum can be
identified with the distance stretched by the magnon in the droplet
plane. In the case of a circular disc, this is given by
$2\sin(p/2)$, whereas for the case of the annulus this is given by
$(\hat{R}_{i+1}-\hat{R}_{i})$.

Considering a BMN-like expansion $\lambda/L^2 \ll 1$ one has
\begin{equation}
\label{esfoldex} E -J -L
=\frac{4\lambda}{L}(\hat{R}_{i+1}-\hat{R}_{i})^{2} +... \,,
\end{equation}
This precisely coincide with the weak coupling field theory result
(\ref{efold}).

It is also interesting to consider the non-static configurations
with a  fast-string limit imposed \cite{Kruczenski,kru}. In this
case the resulting action is
\begin{equation}
\label{grsllm} S = -\frac{L}{2\pi}\int d\tau \int_0^{2\pi}
d\sigma \left(\dot x_a V_a
+\frac{1}{2(\alpha'L)^2}{x_a}'{x_a}'\right)\,,
\end{equation}
In order to facilitate the comparison with field theory results, we
define a normalized complex coordinate in the LLM plane
$z=\sqrt{\frac{\pi}{{{\rm Area}(D)}}}(x_1+i x_2)$. Writing the
LLM functions in complex basis, for concentric droplets one has
$\bar z V(z,\bar z) = - V_\phi(x_1,x_2,0)$ with,
\begin{equation}
V(z,\bar z) = \frac{z}{2 |z|^2} \sum_{k=1}^K (-1)^{K-k}\left(
\frac{|z|^2+{\hat R}_k^2}{||z|^2-{\hat R}_k^2|}-1\right)\,.
\label{ves}
\end{equation}
Here  ${\hat R}_k$ are the normalized radii\footnote{For instance,
for two edges ${\hat R}_1=\sqrt\gamma$ and ${\hat
R}_2=\sqrt{1+\gamma}$.}. Then, the Polyakov action in the
fast-string limit adopts the form
\begin{equation}
\label{rsllm} S = {L}\int d\tau \int_0^{1} d\sigma
\left(\frac{i}{2} V \dot{ \bar z} -\frac{i}{2} \bar V\dot{ z}
-\frac{\lambda}{L^2}|z'|^2\right)\,,
\end{equation}
This again precisely coincides with the coherent state action
(\ref{coac2}) defined for the Hamiltonian (\ref{wbh}).

~

\noindent{\bf Some formulae}

To study the action of the dilatation operator (\ref{d1}) over the
single trace probing the background operator, it is necessary to
consider $\partial_Z$ acting on positive and negative powers of $Z$
considered. For example,
\begin{eqnarray}
&&(\partial_Z)^i_j (Z^n)^k_l = \sum_{a=1}^n
(Z^{a-1})^k_j(Z^{n-a})^i_l\,,
\label{posi}
\\
&&(\partial_Z)^i_j (Z^{-n})^k_l = -\sum_{a=1}^n
(Z^{-a})^k_j(Z^{a-n-1})^i_l\,.
\label{nega}
\end{eqnarray}
A careful inspection shows that, to the leading order in the large
$N$ expansion, the relevant terms correspond to the case where
 an index of $\partial_Z$ is contracted
with an index of $Z^n$. Even in that case, only one term of the sum
contributes
\begin{eqnarray}
&&(\partial_Z)^i_j (Z^n)^j_l = N(Z^{n-1})^i_l + \text{sub-leading}\, ,
\label{posic1}
\\
&&(\partial_Z)^i_j (Z^{n})^k_i = N(Z^{n-1})^k_j + \text{sub-leading}\, .
\label{posic2}
\end{eqnarray}
The subleading terms in (\ref{d1}) give rise to multiple-traces
probing the background operator and are suppressed in the large $N$
limit. The action of $\partial_Z$ is in all other cases always
sub-leading.

In the case of three edges, we need the action of $\partial_Z$
over $P^n$ and $Z_P^n$ and $Z_Q^n$. Again, in the large $N$ limit
the relevant  contributions are
\begin{eqnarray}
&&(\partial_Z)^i_j (P^n)^j_l = N_1(P^{n+1})^i_l + \text{sub-leading}\,
, \label{pn1}
\\
&&(\partial_Z)^i_j (P^n)^k_i = N_1(P^{n+1})^k_j +  \text{sub-leading}\, ,
\label{pn2}
\\
&&(\partial_Z)^i_j (PZ)^j_l = N_1(P)^i_l +  \text{sub-leading}\, ,
\label{zp01}
\\
&&(\partial_Z)^i_j (PZ)^k_i = N_1(P)^k_j + \text{sub-leading}\, ,
\label{zp02}
\\
&&(\partial_Z)^i_j (Z_P^n)^j_l = N(Z_P^{n-1})^i_l
+N_2(Z_Q^{n-1})^i_l +  \text{sub-leading} \quad{\rm for\ } n\geq 1\,,
\label{zpn1}
\\
&&(\partial_Z)^i_j (Z_P^n)^k_i =N(Z_P^{n-1})^k_j
+N_2(Z_Q^{n-1})^k_j +  \text{sub-leading}\quad{\rm for\ } n\geq 1\,,
\label{zpn2}
\\
&&(\partial_Z)^i_j (Z_Q^n)^j_l = N_1(Z_Q^{n-1})^i_l + \text{sub-leading}
\quad{\rm for\ } n\geq 1\,, \label{zqn1}
\\
&&(\partial_Z)^i_j (Z_Q^n)^k_i = N_1(Z_Q^{n-1})^k_j +
\text{sub-leading}\quad{\rm for\ } n\geq 1\,, \label{zqn2}
\end{eqnarray}
%

%%%%%%%%%%%%%%%%%%%%%%%%%%%%%%%%%%%%%%%%%%%%%%%%%%%%%%%%%%%%%%%%%


\begin{thebibliography}{99}

\bibitem{adscft}
J.~M.~Maldacena, ``The large N limit of superconformal field
theories and supergravity,'' Adv.\ Theor.\ Math.\ Phys.\  {\bf 2},
231 (1998) [arXiv:hep-th/9711200].
%%CITATION = HEP-TH 9711200;%%
E.~Witten, ``Anti-de Sitter space and holography,'' Adv.\ Theor.\
Math.\ Phys.\  {\bf 2}, 253 (1998) [arXiv:hep-th/9802150].
%%CITATION = HEP-TH 9802150;%%
S.~S.~Gubser, I.~R.~Klebanov and A.~M.~Polyakov, ``Gauge theory
correlators from non-critical string theory,'' Phys.\ Lett.\ B {\bf
428}, 105 (1998) [arXiv:hep-th/9802109].
%%CITATION = HEP-TH 9802109;%%

\bibitem{Btoy}
D.~Berenstein, ``A toy model for the AdS/CFT correspondence,'' JHEP
{\bf 0407}, 018 (2004) [arXiv:hep-th/0403110].
%%CITATION = HEP-TH 0403110;%%

\bibitem{LLM}
H.~Lin, O.~Lunin and J.~Maldacena, ``Bubbling AdS space and 1/2 BPS
geometries,'' JHEP {\bf 0410} (2004) 025 [arXiv:hep-th/0409174].
%%CITATION = HEP-TH 0409174;%%

\bibitem{Horava}
P.~Horava and P.~G.~Shepard, ``Topology changing transitions in
bubbling geometries,'' JHEP {\bf 0502}, 063 (2005)
[arXiv:hep-th/0502127].
%%CITATION = JHEPA,0502,063;%%

\bibitem{BBPS}
D.~Berenstein, ``Large N BPS states and emergent quantum gravity,''
JHEP {\bf 0601}, 125 (2006) [arXiv:hep-th/0507203].
%%CITATION = HEP-TH 0507203;%%

\bibitem{Brown}
T.~Brown, R.~de Mello Koch, S.~Ramgoolam and N.~Toumbas,
``Correlators, probabilities and topologies in N = 4 SYM,''
arXiv:hep-th/0611290.
%%CITATION = HEP-TH/0611290;%%

\bibitem{sam}
S.~Vazquez, ``Reconstructing 1/2 BPS space-time metrics from matrix
models and spin chains,'' arXiv:hep-th/0612014.
%%CITATION = HEP-TH 0203249;%%

\bibitem{BCV}
D.~Berenstein, D.~H.~Correa and S.~E.~Vazquez, ``Quantizing open
spin chains with variable length: An example from giant gravitons,''
Phys.\ Rev.\ Lett.\  {\bf 95}, 191601 (2005) [arXiv:hep-th/0502172].
%%CITATION = PRLTA,95,191601;%%

\bibitem{BCV2}
D.~Berenstein, D.~H.~Correa and S.~E.~Vazquez, ``A study of open
strings ending on giant gravitons, spin chains and integrability,''
JHEP {\bf 0609}, 065 (2006) [arXiv:hep-th/0604123].
%%CITATION = JHEPA,0609,065;%%


\bibitem{Kruczenski}
M.~Kruczenski, ``Spin chains and string theory,'' Phys.\ Rev.\
Lett.\  {\bf 93}, 161602 (2004) [arXiv:hep-th/0311203].
%%CITATION = HEP-TH 0311203;%%

\bibitem{kru}
M.~Kruczenski, A.~V.~Ryzhov and A.~A.~Tseytlin, ``Large spin limit
of AdS(5) x S**5 string theory and low energy expansion of
ferromagnetic spin chains,'' Nucl.\ Phys.\ B {\bf 692}, 3 (2004)
[arXiv:hep-th/0403120].
%%CITATION = HEP-TH 0403120;%%

\bibitem{minahan}
J.~A.~Minahan and K.~Zarembo, ` `The Bethe-ansatz for N = 4 super
Yang-Mills,'' JHEP {\bf 0303} (2003) 013. [arXiv:hep-th/0212208].
%%CITATION = HEP-TH 0212208;%%


\bibitem{BMN}
D.~Berenstein, J.~M.~Maldacena and H.~Nastase, ``Strings in flat
space and pp waves from N = 4 super Yang Mills,'' JHEP {\bf 0204},
013 (2002) [arXiv:hep-th/0202021].
%%CITATION = HEP-TH 0202021;%%

\bibitem{CJR}
S.~Corley, A.~Jevicki and S.~Ramgoolam, ``Exact correlators of giant
gravitons from dual N = 4 SYM theory,'' Adv.\ Theor.\ Math.\ Phys.\
{\bf 5}, 809 (2002) [arXiv:hep-th/0111222].
%%CITATION = HEP-TH 0111222;%%

\bibitem{zabrodin}
A.~Zabrodin, ``Matrix models and growth processes: From viscous
flows to the quantum  Hall effect,'' arXiv:hep-th/0412219.

\bibitem{Berecotta}
D.~Berenstein and R.~Cotta, ``A Monte-Carlo study of the AdS/CFT
correspondence: an exploration of quantum gravity effects,''
arXiv:hep-th/0702090.

\bibitem{dil}
N.~Beisert, C.~Kristjansen, J.~Plefka, G.~W.~Semenoff and
M.~Staudacher, ``BMN correlators and operator mixing in N = 4 super
Yang-Mills theory,'' Nucl.\ Phys.\  B {\bf 650}, 125 (2003)
[arXiv:hep-th/0208178].
%%CITATION = NUPHA,B650,125;%%

\bibitem{CS}
D.~H.~Correa and G.~A.~Silva, ``Dilatation operator and the super
Yang-Mills duals of open strings on AdS giant gravitons,'' JHEP {\bf
0611}, 059 (2006) [arXiv:hep-th/0608128].
%%CITATION = JHEPA,0611,059;%%

\bibitem{ggas}
R.~de Mello Koch, J.~Smolic and M.~Smolic, ``Giant gravitons - with
strings attached. II,'' arXiv:hep-th/0701067.
%%CITATION = HEP-TH/0701067;%%

\bibitem{GKP2}
 S.~S.~Gubser, I.~R.~Klebanov and A.~M.~Polyakov,
``A semi-classical limit of the gauge/string correspondence,''
Nucl.\ Phys.\  B {\bf 636}, 99 (2002) [arXiv:hep-th/0204051].
%%CITATION = NUPHA,B636,99;%%

\bibitem{ss}
N.~Beisert, J.~A.~Minahan, M.~Staudacher and K.~Zarembo, ``Stringing
spins and spinning strings,'' JHEP {\bf 0309}, 010 (2003)
[arXiv:hep-th/0306139].
  %%CITATION = JHEPA,0309,010;%%

%\cite{Beisert:2005tm}
\bibitem{Beisert:2005tm}
 N.~Beisert,
``The su(2|2) dynamic S-matrix,''
 arXiv:hep-th/0511082.
 %%CITATION = HEP-TH/0511082;%%

\bibitem{CDO2}
H.~Y.~Chen, N.~Dorey and K.~Okamura, ``The asymptotic spectrum of
the N = 4 super Yang-Mills spin chain,'' arXiv:hep-th/0610295.
%%CITATION = HEP-TH/0610295;%%


\bibitem{HM}
D.~M.~Hofman and J.~M.~Maldacena, ``Giant magnons,'' J.\ Phys.\ A
{\bf 39}, 13095 (2006) [arXiv:hep-th/0604135].
%%CITATION = JPAGB,A39,13095;%%

\bibitem{Dorey}
N.~Dorey, ``Magnon bound states and the AdS/CFT correspondence,''
J.\ Phys.\ A  {\bf 39}, 13119 (2006) [arXiv:hep-th/0604175].
%%CITATION = JPAGB,A39,13119;%%

\bibitem{mosaffa}
A.~E.~Mosaffa and M.~M.~Sheikh-Jabbari,
``On classification of the bubbling geometries,''
JHEP {\bf 0604}, 045 (2006)
[arXiv:hep-th/0602270].

\bibitem{Ebrahim}
H.~Ebrahim and A.~E.~Mosaffa, ``Semiclassical string solutions on
1/2 BPS geometries,'' JHEP {\bf 0501}, 050 (2005)
[arXiv:hep-th/0501072].
%%CITATION = JHEPA,0501,050;%%


\bibitem{pp}
Y.~Takayama and K.~Yoshida, ``Bubbling 1/2 BPS geometries and
Penrose limits,'' Phys.\ Rev.\  D {\bf 72}, 066014 (2005)
[arXiv:hep-th/0503057].
%%CITATION = PHRVA,D72,066014;%%

%\cite{Blau:2001ne}
\bibitem{blau}
M.~Blau, J.~Figueroa-O'Farrill, C.~Hull and G.~Papadopoulos, ``A new
maximally supersymmetric background of IIB superstring theory,''
JHEP {\bf 0201}, 047 (2002) [arXiv:hep-th/0110242].
%%CITATION = JHEPA,0201,047;%%

\bibitem{met}
R.~R.~Metsaev, ``Type IIB Green-Schwarz superstring in plane wave
Ramond-Ramond background,'' Nucl.\ Phys.\  B {\bf 625}, 70 (2002)
[arXiv:hep-th/0112044].

\bibitem{Filev}
V.~Filev and C.~V.~Johnson, ``Operators with large quantum numbers,
spinning strings, and giant gravitons,'' Phys.\ Rev.\  D {\bf 71},
106007 (2005) [arXiv:hep-th/0411023].
%%CITATION = PHRVA,D71,106007;%%


\bibitem{Alishahiha}
M.~Alishahiha, H.~Ebrahim, B.~Safarzadeh and M.~M.~Sheikh-Jabbari,
``Semi-classical probe strings on giant gravitons backgrounds,''
JHEP {\bf 0511}, 005 (2005) [arXiv:hep-th/0509160].
%%CITATION = JHEPA,0511,005;%%

\bibitem{Aru}
G.~Arutyunov and S.~Frolov, ``Integrable Hamiltonian for classical
strings on AdS(5) x S**5,'' JHEP {\bf 0502}, 059 (2005)
[arXiv:hep-th/0411089].
%%CITATION = JHEPA,0502,059;%%

\end{thebibliography}
\end{document}